# A Simple Characterization of Strategic Behaviors in Broadcast Channels

Yi Su[*] and Mihaela van der Schaar

*Abstract* — In this paper, we consider the problem of resource allocation among two competing users sharing a binary symmetric broadcast channel. We model the interaction between autonomous selfish users in the resource allocation and analyze their strategic behavior in manipulating the allocation outcome. We analytically show that users will improve their performance (i.e. gain higher allocated rates) if they have more information about the strategy of the competing user.

## I. Introduction

Game theory has recently been applied to model and characterize the multi-user interaction in communication settings such as the information-theoretic multi-access channels [1], interference channels [2][3] and operational contention-based random access channels [4]. Issues such as fairness for multi-user interactions in allocating rates to users have been studied [1][2]. The existence and performance of equilibrium are studied for multi-user frequency-selective interference channels and ALOHA networks [3][4].

As opposed to prior work, in which the main goal is mostly to examine the system performance of the wireless system in a game theoretic setting from a system planner perspective, in this letter we focus on studying the strategic behavior of the selfish users, i.e. their ability in manipulating the resource allocation outcome. We study this problem in a simple setting: the binary symmetric broadcast channels shared by two competing users. We model the interaction in the resource allocation among autonomous selfish users and analyze their behavior. We further consider the cases in which one user has different amounts of information about the strategy of other user. We show that this information benefits strategic users and results in an improved utility. While our results aim at quantifying the benefits of having additional information about competing users in a simple two-user setting, they motivate the need for additional research in studying the multi-user strategic interaction in the wireless environment.

## II. THE TWO-USER INTERACTION FOR BINARY SYMMETRIC BROADCAST CHANNELS

We consider a broadcast channel consisting of a pair of binary symmetric channels (BSC) with parameters (i.e. the cross-over probabilities) $p_1$ and $p_2$ (see pp. 568-570 [5]). Without loss of generality, we assume in this letter that $0 < p_1 < p_2 < 0.5$. The capacity region for this channel is the convex hull of the closure of all $(R_1, R_2)$ satisfying

$$\begin{aligned} R_1 &\leq H(\beta * p_1) - H(p_1) \\ R_2 &\leq 1 - H(\beta * p_2), \end{aligned} \quad (1)$$

where $H(x) = -x\log x - (1-x)\log(1-x)$, $\beta * p = \beta(1-p) + (1-\beta)p$, $\beta \in [0, 0.5]$ [5]. It can be easily shown that the convex hull operation is not necessary. Therefore, the Pareto surface of the capacity region is given by

$$\begin{aligned} R_1 &= H(\beta * p_1) - H(p_1) \\ R_2 &= 1 - H(\beta * p_2). \end{aligned} \quad (2)$$

There are two *users* and a *resource manager* in the system and each user occupies a BSC. We denote by $N = \{1, 2\}$ the set of users. User $i$ holds the *private information* about its BSC parameter $p_i$, which lies in the set $\Theta_i = [0, 0.5]$ [5]. User $i$ announces the *public information* $p'_i$ about its $p_i$. Note that for strategic users, $p'_i$ can differ from $p_i$. For user $i \in N$, the set of *actions* available for user $i$ to announce $p'_i$ is denoted $A_i$ and $A_i = [0, 0.5]$. Let $\boldsymbol{p}' = (p'_1, p'_2)$ and $A = A_1 \times A_2$. Users have preferences over the outcomes of the resource allocation and these preferences are represented by a *utility function*, $u_i : A \times \Theta_i \to \mathcal{R}$, where $u_i$ is the actual rate $R_i$ that user $i$ can achieve in the allocation outcome. Fig. 1 shows the diagram of the resource allocation procedure. The resource manager calculates user $i$'s achievable rate $R'_i$ based on the public information $\boldsymbol{p}'$ and allocates the resources based on certain policy $\mathcal{P}$ that indicates the system goal. In this letter, we assume that the resource is divided based on the well-known proportional fair allocation [6], i.e. $R'_1 R'_2$ is maximized, but other allocation rules could also be implemented and they will result in different strategic behavior. Summarizing, the tuple $\langle N, (\Theta_i), (A_i), (u_i), \mathcal{P} \rangle$ defines the model of interaction between the users [7].

The resource manager collects the public information $\boldsymbol{p}'$ and determines the optimal $\beta$ that maximizes $R'_1 R'_2$. Because the capacity region is convex, the optimal value uniquely exists. Taking the derivative with respect to $\beta$, we

have

$$\frac{\partial \{R_1'R_2'\}}{\partial \beta} = [1 - H(\beta * p_2')](1 - 2p_1')\log\frac{1 - \beta * p_1'}{\beta * p_1'} - [H(\beta * p_1') - H(p_1')](1 - 2p_2')\log\frac{1 - \beta * p_2'}{\beta * p_2'}. \quad (3)$$

The optimum $\beta = \beta_{opt}$ satisfies $\left.\frac{\partial \{R_1'R_2'\}}{\partial \beta}\right|_{\beta=\beta_{opt}} = 0$, $0 \leq \beta \leq \frac{1}{2}$. Note that $\beta_{opt}$, $u_i$, and $R_i$ are functions of $p'$. We will use $\beta_{opt}$, $u_i$, $R_i$, $\beta_{opt}(p_1', p_2')$, $u_i(p_1', p_2')$, and $R_i(p_1', p_2')$ interchangeably hereafter. After the manager determines $\beta_{opt}$, users get their utilities. For example, if $p_1' < p_2'$, $u_1 = R_1 = H(\beta_{opt} * p_1) - H(p_1)$ and $u_2 = R_2 = 1 - H(\beta_{opt} * p_2)$. Note that in this letter we focus on the allocation outcome in the information theoretic sense, rather than operational issues.

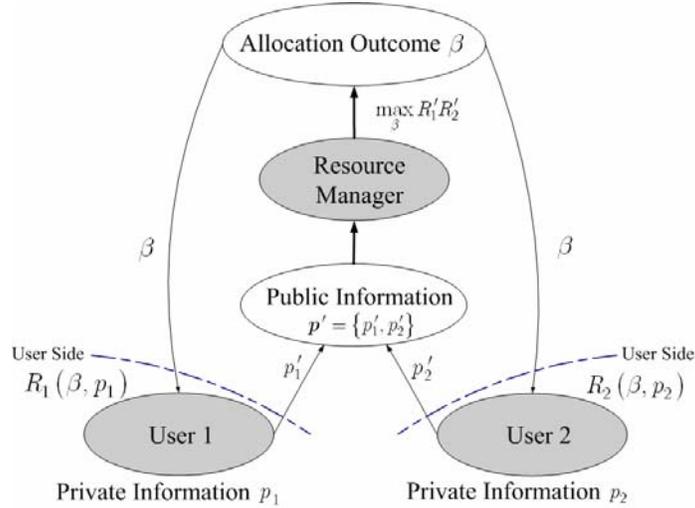

Fig. 1. System diagram.

Existing work in the broadcast channel usually assumes that the resource allocation is performed based on the channel state information (CSI) fed back from the receivers. Since the system goal does not always coincide with the goals of the selfish users, the users have the incentives to strategically change their reported CSI, i.e. the announced public information $p'$, to benefit themselves. Each selfish user plays strategically to maximize its own utility $u_i$ by announcing appropriate public information $p_i' \in A_i$, i.e.

$$\max_{p_i' \in A_i} R_i. \quad (4)$$

## III. BEST RESPONSE IN THE TWO-USER INTERACTION

In order to understand the behavior of individual users, here we first assume that user 1 reports $p_1' = \theta$ and user 2 strategically reports $p_2'$ instead of $p_2$. The resource manager will treat the cases in which $p_2' \geq \theta$ and $p_2' < \theta$ separately by transforming into different physically degraded channels [5]. To determine the optimal value of $p_2'$ in (4), we

discuss two cases:

- First, if $p'_2 \geq \theta$, the rate user 2 can achieve is $R_2 = 1 - H(\beta_{opt} * p_2)$. Therefore, the problem in (4) is identical to
$$\min_{p'_2 \in A_2} \beta_{opt}(\theta, p'_2). \tag{5}$$

- Second, if $p'_2 < \theta$, the rate user 2 can achieve now is $R_2 = H(\beta_{opt} * p_2) - H(p_2)$ and problem (4) is identical to
$$\max_{p'_2 \in A_2} \beta_{opt}(p'_2, \theta). \tag{6}$$

*Lemma 1:* The best strategy for user 2 to maximize $R_2$ is to report its public information as $p'_2 = p'_1 = \theta$.

*Proof:* To show that $p'_2 = p'_1 = \theta$ maximizes $R_2$, we need to demonstrate that $\dfrac{\partial \beta_{opt}}{\partial p'_2} > 0$ for both $p'_2 < \theta$ and $p'_2 \geq \theta$.

First, we consider the case with $p'_2 < \theta$. Denote $g(p'_1, p'_2, \beta) = \dfrac{R'_1}{R'_2} \dfrac{\partial R'_2}{\partial R'_1} = -\dfrac{H(\beta * p'_1) - H(p'_1)}{1 - H(\beta * p'_2)} \cdot \dfrac{1 - 2p'_2}{1 - 2p'_1} \cdot \dfrac{\log \dfrac{1 - \beta * p'_2}{\beta * p'_2}}{\log \dfrac{1 - \beta * p'_1}{\beta * p'_1}}$.

The resource manager always chooses to operate at $\beta = \beta_{opt}$ with $R'_1 \dfrac{\partial R'_2}{\partial \beta}\bigg|_{\beta = \beta_{opt}} + R'_2 \dfrac{\partial R'_1}{\partial \beta}\bigg|_{\beta = \beta_{opt}} = 0$, i.e.

$g(p'_1, p'_2, \beta_{opt}) = -1$. In the following, we will show that $\dfrac{\partial g(p'_1, p'_2, \beta)}{\partial p'_1} > 0$, $\dfrac{\partial g(p'_1, p'_2, \beta)}{\partial p'_2} > 0$, and $\dfrac{\partial g(p'_1, p'_2, \beta)}{\partial \beta} < 0$.

Because of these properties, if $p'_2$ increases, $\beta_{opt}$ needs to be increased in order to keep $g(p'_1, p'_2, \beta_{opt}) = -1$. Hence, we can conclude that $\dfrac{\partial \beta_{opt}}{\partial p'_2} > 0$. Using a similar argument, we can conclude that $\dfrac{\partial \beta_{opt}}{\partial p'_1} > 0$.

To show $\dfrac{\partial g(p'_1, p'_2, \beta)}{\partial p'_1} > 0$, $\dfrac{\partial g(p'_1, p'_2, \beta)}{\partial p'_2} > 0$, and $\dfrac{\partial g(p'_1, p'_2, \beta)}{\partial \beta} < 0$, we have

$$g(p'_1, p'_2, \beta) = C_1 \dfrac{H(\beta * p'_1) - H(p'_1)}{(1 - 2p'_1) \log \dfrac{1 - \beta * p'_1}{\beta * p'_1}} = C_2 \dfrac{(1 - 2p'_2) \log \dfrac{1 - \beta * p'_2}{\beta * p'_2}}{1 - H(\beta * p'_2)} = C_3 \dfrac{(H(\beta * p'_1) - H(p'_1)) \log \dfrac{1 - \beta * p'_2}{\beta * p'_2}}{(1 - H(\beta * p'_2)) \log \dfrac{1 - \beta * p'_1}{\beta * p'_1}}, \tag{7}$$

in which $C_1 = -\dfrac{(1 - 2p'_2) \log \dfrac{1 - \beta * p'_2}{\beta * p'_2}}{1 - H(\beta * p_2)}$, $C_2 = -\dfrac{H(\beta * p'_1) - H(p'_1)}{(1 - 2p'_1) \log \dfrac{1 - \beta * p'_1}{\beta * p'_1}}$, and $C_3 = -\dfrac{1 - 2p'_2}{1 - 2p'_1}$.

Denote $D_1 = (1 - 2p'_1)^{-2} \left[\log \dfrac{1 - \beta * p'_1}{\beta * p'_1}\right]^{-2}$, $D_2 = [1 - H(\beta * p'_2)]^{-2}$, and $D_3 = \left[(1 - H(\beta * p'_2)) \log \dfrac{1 - \beta * p'_1}{\beta * p'_1}\right]^{-2}$.

Taking the derivative with respect to $p'_1$, we have

$$\frac{\partial g(p_1', p_2', \beta)}{\partial p_1'} = C_1 D_1 \left\{ (1 - 2p_1')(1 - 2\beta) \left[ \left( \log \frac{1 - \beta * p_1'}{\beta * p_1'} \right)^2 + \frac{H(\beta * p_1') - H(p_1')}{(\beta * p_1')(1 - \beta * p_1')} \right] \right.$$
$$\left. - \log \frac{1 - \beta * p_1'}{\beta * p_1'} \left[ (1 - 2p_1') \log \frac{1 - p_1'}{p_1'} - 2(H(\beta * p_1') - H(p_1')) \right] \right\}. \tag{8}$$

Clearly, $\frac{\partial g(p_1', p_2', \beta)}{\partial p_1'} > 0$ because

$$2(H(\beta * p_1') - H(p_1')) < \frac{\partial H(p_1')}{\partial p_1'}(\beta * p_1' - p_1') = 2\beta(1 - 2p_1') \log \frac{1 - p_1'}{p_1'} < (1 - 2p_1') \log \frac{1 - p_1'}{p_1'}.$$

Taking the derivative with respect to $p_2'$, we have

$$\frac{\partial g(p_1', p_2', \beta)}{\partial p_2'} = C_2 D_2 \left\{ (1 - 2\beta)(1 - 2p_2') \left[ -\frac{1 - H(\beta * p_2')}{(\beta * p_2')(1 - \beta * p_2')} - \left( \log \frac{1 - \beta * p_2'}{\beta * p_2'} \right)^2 \right] \right.$$
$$\left. - 2 \log \frac{1 - \beta * p_2'}{\beta * p_2'} \left[ 1 - H(\beta * p_2') \right] \right\} > 0. \tag{9}$$

Taking the derivative with respect to $\beta$, we have

$$\frac{\partial g(p_1', p_2', \beta)}{\partial \beta} = C_3 D_3 \left\{ (1 - H(\beta * p_2'))(H(\beta * p_1') - H(p_1')) \left[ \frac{1 - 2p_1'}{(1 - \beta * p_1')(\beta * p_1')} \cdot \log \frac{1 - \beta * p_2'}{\beta * p_2'} \right. \right.$$
$$\left. - \frac{(1 - 2p_2')}{(\beta * p_2')(1 - \beta * p_2')} \cdot \log \frac{1 - \beta * p_1'}{\beta * p_1'} \right] + \log \frac{1 - \beta * p_2'}{\beta * p_2'} \cdot \log \frac{1 - \beta * p_1'}{\beta * p_1'} \cdot \tag{10}$$
$$\left[ (1 - H(\beta * p_2')) \log \frac{1 - \beta * p_1'}{\beta * p_1'}(1 - 2p_1') + (H(\beta * p_1') - H(p_1')) \log \frac{1 - \beta * p_2'}{\beta * p_2'}(1 - 2p_2') \right].$$

Let $f(\beta, p) = \frac{1 - 2p}{(1 - \beta * p)(\beta * p) \log \frac{1 - \beta * p}{\beta * p}}$. It is easy to show that $\frac{\partial f(\beta, p)}{\partial p} < 0$. Therefore, we have

$f(\beta, p_2') > f(\beta, p_1')$, which leads to $\frac{\partial g(p_1', p_2', \beta)}{\partial \beta} < 0$.

Second, we consider the case in which $p_2' < \theta$. Using the fact that $\frac{\partial \beta_{opt}}{\partial p_1'} > 0$ for $p_2' < p_1'$, by symmetry, we can also conclude that $\frac{\partial \beta_{opt}}{\partial p_2'} > 0$ for $p_2' \geq p_1'$.

Therefore, we have $\frac{\partial \beta_{opt}}{\partial p_2'} > 0$ for both $p_2' < p_1'$ and $p_2' \geq p_1'$. The optimal solution of (4) is $p_2' = p_1' = \theta$. ∎

We simulate an example in which $p_1 = p_1' = 0.1$ and $p_2 = 0.2$. The result is shown in Fig. 2 and 3. We can see that $\beta_{opt}$ monotonically increases in $p_2'$ and the best strategy in maximizing $R_2$ is to let $p_2' = p_1' = 0.1$.

In fact, given $p_1' = \theta$, considering the discontinuity of the allocated rate $R_2$ as a function of $p_2'$ (see Fig. 3), user 2 will always choose to announce that $p_2' = \theta + \varepsilon$, where $\varepsilon > 0$ and $\varepsilon \to 0$. Similarly, given $p_2' = \xi$, user 1 will always choose to announce that $p_1' = \xi - \varepsilon$. Therefore, for $p_1' = \theta$ and $p_2' = \theta + \varepsilon$, where $\varepsilon > 0$ and $\varepsilon \to 0$, nobody will deviate from this point because no user can gain more utility by unilaterally changing its action $p'_i \in A_i$. Hence, based on the definition of the Nash equilibrium [7], we can conclude that the considered interaction between the users reaches Nash equilibrium.

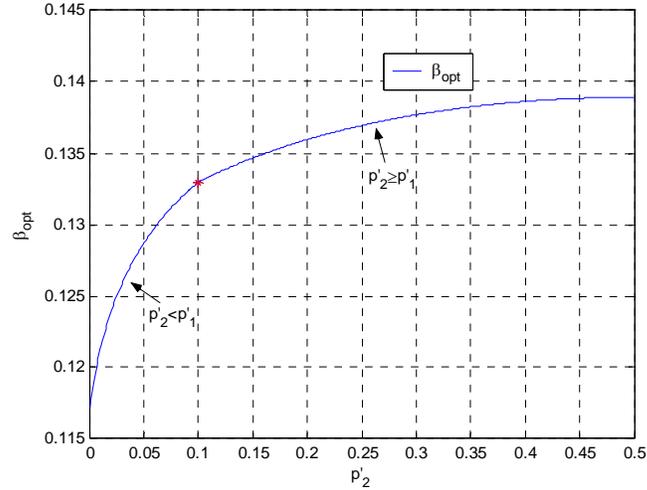

Fig. 2. $\beta_{opt}$ for different announced $p'_2$.

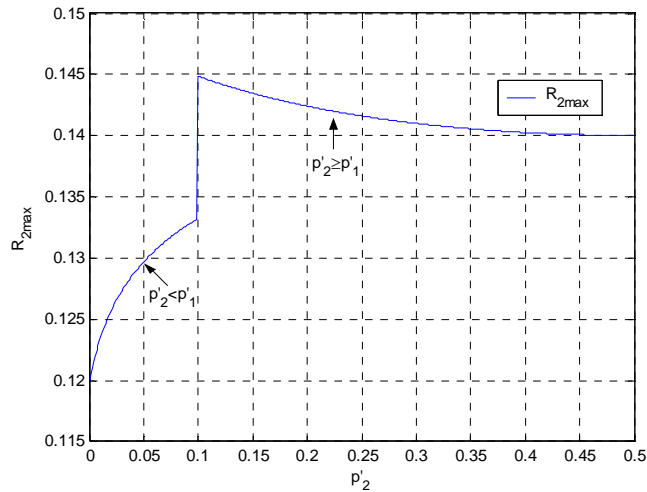

Fig. 3. Actual allocated rate $R_2$ for different announced $p'_2$.

For an intuitive explanation about the result, we resort to the basic idea of the superposition coding for the degraded broadcast channel (Theorem 15.6.2, pp.565-567, [5]). An auxiliary random variable will serve as a cloud center distinguishable by both receivers. Each cloud consists of $2^{nR_1}$ codewords. The worst receiver can only see the clouds, while the better receiver can see the individual codewords within the clouds. The worse receiver will always want to decrease the size of the clouds, since it can only distinguish the clouds and decreasing the size of clouds will increase his achievable rate. On the contrary, the better receiver will always want to increase the distance between the centers of the clouds so that he can accommodate more bits in each cloud. In other words, the better receiver wants to increase the size of clouds. Specifically, in our problem, $\beta$ is the variable indicating the size of the clouds. Recall that $\beta = 0$ when $R_1 = 0$ (i.e. cloud size is zero) and $\beta = 0.5$ when $R_2 = 0$ (i.e. cloud is large and overlapped such that user 2 gets nothing). In short, the worse receiver wants a small $\beta$ but the better receiver prefers a large $\beta$. In Lemma 1, we have already shown that $\frac{\partial \beta_{opt}}{\partial p'_2} > 0$. By "moderately bragging", a worse receiver can decrease the size of the clouds and increases his rate. However, if he excessively brags, he will become "the better receiver" conceptually and he needs to decode within each cloud so that shrinking the cloud size does no benefit him. The discontinuity comes from the fact that if $p'_2$ goes across $p_1$, the roles of the better receiver and the worse receiver switch, which causes the achievable rates of these two different coding schemes not to be continuous at the boundary of $p'_2 = p_1$ unless $p_2 = p_1$.

## IV. IMPACT OF SIDE INFORMATION

In this section, based on the results of the previous section, we briefly discuss the impacts of the side information about the other user's strategy. We will show that a user having additional information about the strategy of the other user will increase its utility. In practice, it is usually difficult for users to accurately access the complete information about the other user's strategy, but it is possible to have some side information about them. This kind of side information could be obtained by repeatedly participating in the resource allocation and observing the multi-user interaction. We assume that the public information $p'_1$ that user 1 announces is a random variable $Y$ with the probability density function (pdf) $p'_1(y)$, $y \in [0, 0.5]$. User 2 has side information $Z$ relevant to $p'_1$ with pdf $p(z), z \in (-\infty, +\infty)$. Assume $Y$ and $Z$ have joint pdf $p'_1(y,z) = p(z) p'_1(y \mid z)$, $y \in [0, 0.5]$, $z \in (-\infty, +\infty)$. User 2 determines its response $p'_2 \mid z$ such that its utility, the expectation of its allocated rate $E_{p_1(y,z)}[R_2]$, is maximized:

$$\max_{p'_2 \mid z \in A_2} E_{p_1(y,z)}[R_2]. \tag{11}$$

$E_{p_1(y,z)}[R_2]$ can be expressed as

$$E_{p_1(y,z)}[R_2] = \int_{-\infty}^{+\infty} \int_0^{0.5} p'_1(y,z) R_2(y, p'_2 \mid z) dy dz = \int_{-\infty}^{+\infty} p(z) \cdot \left( \int_0^{0.5} p'_1(y \mid z) R_2(y, p'_2 \mid z) dy \right) dz, \tag{12}$$

where $R_2(y, p'_2 \mid z)$ represents the rate allocated to user 2 with public information $y$ and $p'_2 \mid z$.

Therefore, this expectation-maximization strategy is equivalent to

$$\max_{p'_2|z \in A_2} \int_0^{0.5} p'_1(y \mid z) R_2(y, p'_2 \mid z) dy. \tag{13}$$

Now we consider three types of $p'_1(y \mid z)$ with different amounts of side information and examine the corresponding impacts. In the first case, suppose there exists an injective function $f : [0, 0.5] \to (-\infty, +\infty)$ and $z = f(y)$. Hence, $Z$ provides all the information about $Y$ and $p'_1(y \mid z) = \delta(y - f^{-1}(z))$. Second, we consider the general $p'_1(y \mid z)$ as an intermediate case, in which no injective function mapping $Y$ to $Z$ exists and $p'_1(y \mid z) \neq p'_1(y)$. In the third case, $Z$ is independent of $Y$, i.e. $Z$ provides no information about $Y$ and $p'_1(y \mid z) = p'_1(y)$. The best response and corresponding expected utility $R_2^i$ that user 2 can achieve in the $i$th case is different. In the following, we will investigate these different cases in detail.

In the first case, given the side information $z$, from Lemma 1, the best strategy is to announce $p'_2 = f^{-1}(z)$. Thus,

$$\begin{aligned}
R_2^1 &= \int_{-\infty}^{+\infty} p(z) \cdot \left( \max_{p'_2|z \in A_2} \int_0^{0.5} p'_1(y \mid z) R_2(y, p'_2 \mid z) dy \right) dz \\
&= \int_{-\infty}^{+\infty} p(z) \cdot \left( \max_{p'_2|z \in A_2} \int_0^{0.5} \delta(y - f^{-1}(z)) R_2(y, p'_2 \mid z) dy \right) dz \\
&= \int_0^{0.5} p'_1(y) R_2(y, y) dy.
\end{aligned} \tag{14}$$

In the second case, $R_2^2$ is given in (12) for the general $p'_1(y \mid z)$.

In the third case, we have $p'_1(y, z) = p(z) p'_1(y \mid z) = p(z) p'_1(y)$, and therefore

$$\begin{aligned}
R_2^3 &= \int_{-\infty}^{+\infty} p(z) \cdot \left( \max_{p'_2|z \in A_2} \int_0^{0.5} p'_1(y \mid z) R_2(y, p'_2 \mid z) dy \right) dz \\
&= \int_{-\infty}^{+\infty} p(z) \cdot \left( \max_{p'_2|z \in A_2} \int_0^{0.5} p'_1(y) R_2(y, p'_2 \mid z) dy \right) dz \\
&= \max_{p'_2 \in A_2} \int_0^{0.5} p'_1(y) R_2(y, p'_2) dy.
\end{aligned} \tag{15}$$

The second equality holds because the maximizer of $p'_2 \mid z$ is independent of $z$.

Denote $p'^*_2 \mid z = \arg\max_{p'_2|z \in A_2} \int_0^{0.5} p'_1(y \mid z) R_2(y, p'_2 \mid z) dy$ and $p'^*_2 = \arg\max_{p'_2 \in A_2} \int_0^{0.5} p'_1(y) R_2(y, p'_2) dy$. By Lemma 1, we know that $R_2(y, y) > R_2(y, p'_2), \forall p'_2 \neq y$. Therefore, in the second case, $\exists z \in (-\infty, +\infty), [p'^*_2 \mid z] \neq y, R_2(y, p'^*_2 \mid z) < R_2(y, y)$, otherwise we can construct an injective function that maps $Y$ to $Z$. Consequently, we have

$$R_2^2 = \int_{-\infty}^{+\infty} p(z) \cdot \left[ \max_{p_2'|z \in A_2} \int_0^{0.5} p_1'(y|z) R_2(y, p_2'|z) dy \right] dz$$
$$= \int_0^{0.5} p_1'(y) \int_{-\infty}^{+\infty} p_1'(z|y) R_2\left[y, \left(p_2'^*|z\right)\right] dz dy$$
$$< \int_0^{0.5} p_1'(y) \int_{-\infty}^{+\infty} p_1'(z|y) R_2(y,y) dz dy \qquad (16)$$
$$= \int_0^{0.5} p_1'(y) R_2(y,y) dy = R_2^1,$$

By setting $p_2'^*|z = p_2'^*$, we can show another inequality

$$R_2^2 = \int_{-\infty}^{+\infty} p(z) \cdot \left[ \max_{p_2'|z \in A_2} \int_0^{0.5} p_1'(y|z) R_2(y, p_2'|z) dy \right] dz$$
$$> \int_{-\infty}^{+\infty} p(z) \cdot \left( \int_0^{0.5} p_1'(y|z) R_2\left(y, p_2'^*\right) dy \right) dz \qquad (17)$$
$$= \int_0^{0.5} p_1'(y) R_2\left(y, p_2'^*\right) dy = R_2^3.$$

The inequality in (17) strictly holds, because in the second case $p_1'(y|z) \neq p_1'(y)$, $\exists z \in (-\infty, +\infty)$, $\left[p_2'^*|z\right] \neq p_2'^*$, and $\int_0^{0.5} p_1'(y|z) R_2\left(y, p_2'^*|z\right) dy > \int_0^{0.5} p_1'(y|z) R_2\left(y, p_2'^*\right) dy$.

Combining the two inequalities, we have

$$R_2^1 > R_2^2 > R_2^3. \qquad (18)$$

From (18), we can conclude that more side information about the strategy of the others will help the strategic users improve their performance. Therefore, strategic users in the resource allocation have the incentive to obtain as much side information about the strategy of other users as possible. To quantify the amount of this side information, we can define it based on the achieved utility:

$$I_U(Y;Z) = \int_{-\infty}^{+\infty} p(z) \int_0^{0.5} p_1'(y|z) \left\{ u_2\left[y, \left(p_2'^*|z\right)\right] - u_2\left(y, p_2'^*\right) \right\} dy dz$$
$$= \int_{-\infty}^{+\infty} p(z) \int_0^{0.5} p_1'(y|z) \left\{ R_2\left[y, \left(p_2'^*|z\right)\right] - R_2\left(y, p_2'^*\right) \right\} dy dz \qquad (19)$$

Note that $p_2'^*|z = \arg \max_{p_2'|z \in A_2} \int_0^{0.5} p_1'(y|z) R_2(y, p_2'|z) dy$, which is a function of $p_1'(y|z)$. Therefore, $u_2\left[y, \left(p_2'^*|z\right)\right]$ is a function of $p_1'(y|z)$. Similarly, because $p_2'^* = \arg \max_{p_2' \in A_2} \int_0^{0.5} p_1'(y) R_2(y, p_2') dy$, $u_2\left(y, p_2'^*\right)$ is a function of $p_1'(y)$. We can see that $I_U(Y;Z)$ coincides with the well-known mutual information $I(Y;Z)$ [5] if the utility functions take the form

$$\begin{aligned} u_2\left[y,\left(p_2'^*\mid z\right)\right] &= \log\left(p_1'(y\mid z)\right) \\ u_2\left(y, p_2'^*\right) &= \log\left(p_1'(y)\right) \end{aligned}. \quad (20)$$

In other words, the mutual information $I(Y;Z)$ is a special case of this utility-based information measure $I_U(Y;Z)$. Strategic users always try to obtain as much amount of $I_U(Y;Z)$ as possible in order to benefit themselves. In practice, this side information could be attained if the users are equipped with advanced radios, such as cognitive radios, with the abilities to sense, infer, and learn the environment.

## V. Conclusions

In this letter, we study the strategic behavior by presenting an example of a two-user binary symmetric broadcast channel. We examine the interactions between autonomous selfish users in the resource allocation and analyze their strategic behaviors. We also explicitly show that users will improve their performance if they have additional information about the strategy of other users. Hence, although in this work we only consider a simple two-user binary symmetric broadcast channel and assume the rule of proportional fairness, our results show that determining how to attain this additional information in a multi-user strategic interaction setting deserves investigation in future research.